# Quantum Mechanics Interpreted Through Quantum Electrodynamics

Eduardo V. Flores
Physics & Astronomy
Rowan University
Glassboro, NJ 08028
flores@rowan.edu

We analyze an experiment in which a thin wire is scanned across the overlap of two in-phase photon beams. We find that unless the wire induces the formation of an interference pattern, the complementarity inequality is violated. Quantum electrodynamics accounts for this effect through photon deflections mediated by energy–momentum exchange with the wire via virtual particles. This mechanism underscores the essential role of virtual particles in producing field effects and motivates an interpretation of quantum mechanics in which the wavefunction or field serves as information—a blueprint guiding virtual particles, whose exchanges with external sources shape the trajectories of real particles in accordance with field predictions. As an application, we propose a test of gravity's quantum nature.

Classical electrodynamics provides a robust framework for describing many properties of light [1]. For example, when two uniform, in-phase light beams intersect, an interference pattern forms on a screen placed in the region of overlap. According to classical theory, regions of destructive interference correspond to zero electromagnetic field intensity and are therefore assumed to contain no photons. Conversely, in regions of constructive interference, the electromagnetic fields reinforce each other and photon density increases. This pattern is expected to exist even in the absence of a detection screen. However, as we show here, this classical interpretation of interference can lead to a violation of the principle of complementarity.

To illustrate this, we consider an experimental setup in which coherent light with a wavelength of 632 nm is incident on a 50:50 beam splitter (Fig. 1). The beam splitter produces two statistically independent beams [2]. These beams intersect at a small angle and then separate, ultimately reaching their respective detectors [3]. The experiment can be performed under conditions of extremely low photon flux, such that at most one photon is present in the entire apparatus at any given time [3v4]. Alternatively, it can be conducted using true single-photon sources.

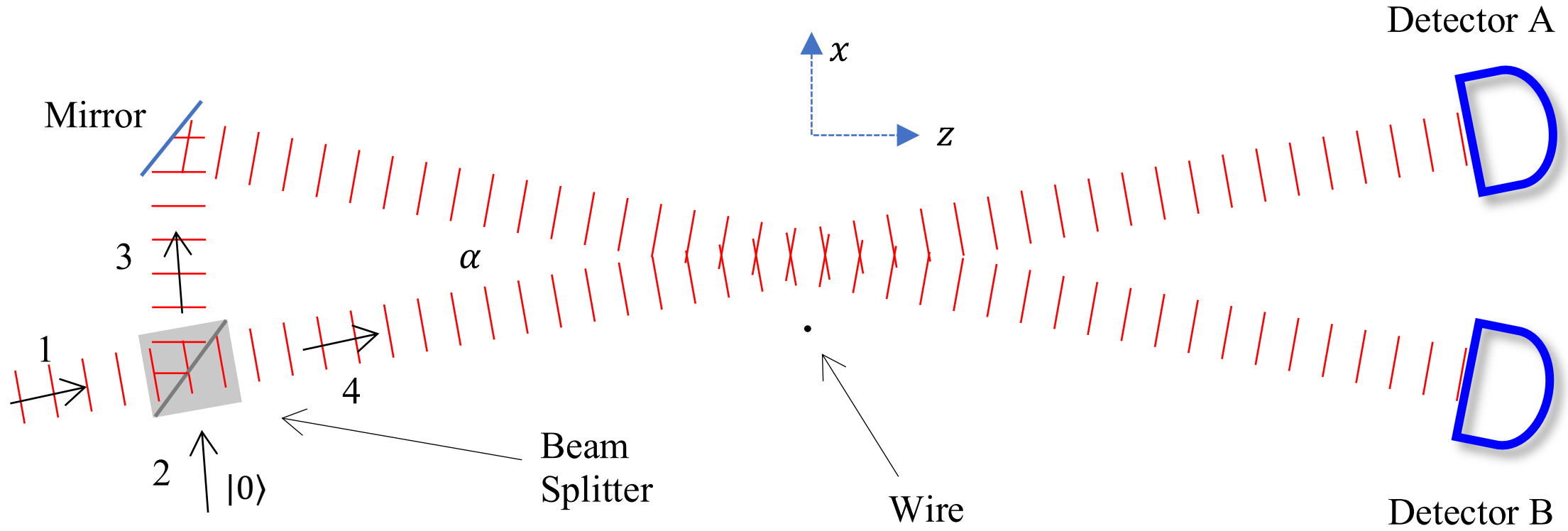

**Fig. 1 | Experimental setup with identical in-phase laser beams.**
A laser beam enters arm 1 of a beam splitter, while arm 2 remains empty. The beams in output arms 3 and 4 cross at an angle $\alpha = 2.971$ mrad and impinge on separate detectors. A 17 $\mu$m–thick wire, aligned parallel to the intensity fringes, is scanned across the 1.0 mm beam intersection to probe the interference pattern. The distance from the beam splitter to the wire is 0.454 m, and from the wire to the detectors is 2.521 m.

When two in-phase plane waves originate from separate locations, as illustrated in Fig. 1, they interfere to produce a spatially varying electric field intensity given by

$$E_0^2(1 + \cos[2\pi\alpha x/\lambda]), \tag{1}$$

where $x$ is the transverse coordinate across the beam intersection, $\alpha$ is the angle between the beams, and $\lambda$ is the wavelength of the light. If the beams terminate on a screen placed at the beam intersection, interference fringes with perfect visibility ($V = 1$) are observed. In this case, we cannot determine whether a photon originated at the mirror or at the beam splitter, and the path information is zero ($K = 0$).

Conversely, when the screen is removed, a click at detector A indicates that the photon came from the beam splitter, while a click at detector B indicates that it came from the mirror. Thus, as the photon crosses the beam intersection, its path is known with certainty ($K = 1$), but the visibility is zero ($V = 0$). All of this is consistent with the principle of complementarity [4,5]:

$$K^2 + V^2 \leq 1. \tag{2}$$

Using the setup shown in Fig. 1, it is possible to simultaneously measure visibility ($V$) and path information ($K$). The technique involves placing a thin wire parallel to the interference fringes [6]. In our case, a 17 $\mu$m –thick wire is scanned across the beam intersection [3]. This wire is considerably thinner than the fringe spacing, which is approximately 212 $\mu$m.

We define the ratio $f$ as the photon count with the wire in place divided by the photon count without the wire. This ratio is plotted as a function of the wire position in Fig. 2. The solid line represents the theoretical prediction for $f$, based on a model that combines Babinet's principle with classical wire diffraction [1,3]. The theoretical calculations are in close agreement with the experimental data.

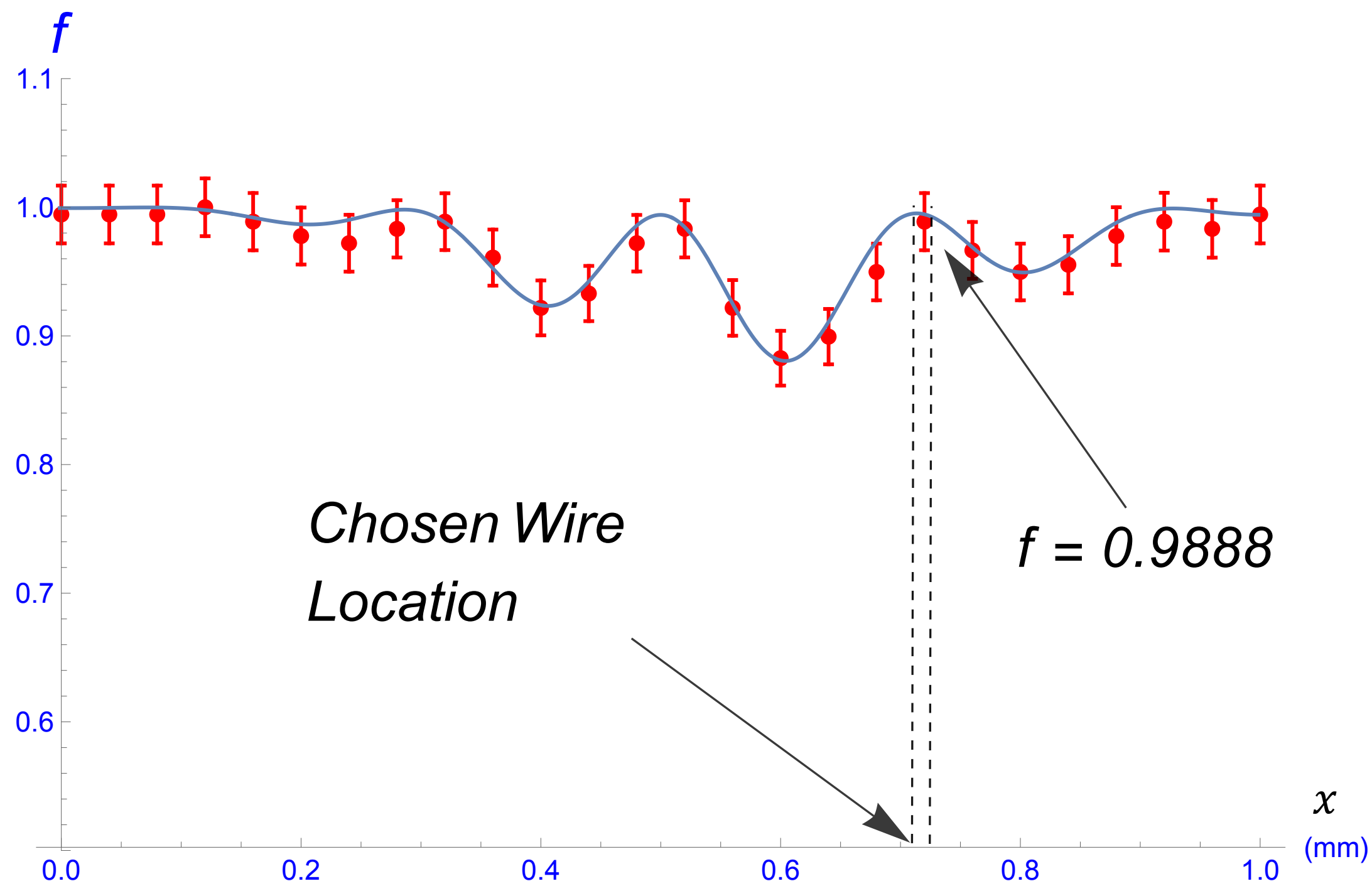

**Fig. 2 | Fraction of photon count $f$ at the end detector.**
Measured fraction of photon count at one of the end detectors as a 17 $\mu$m –thick wire is scanned across a 1.0 mm–wide beam intersection. Error bars indicate statistical uncertainties in the photon count. The solid curve shows the theoretical prediction. Notably, the pattern remains unchanged even when the average separation between successive photons corresponds to a distance of 3 km.

From Fig. 2, we observe that the fraction of photon count $f$ decreases to 0.8827 when the wire is positioned at $x = 0.60$ mm, indicating the presence of a bright interference fringe at that location. This reduction corresponds to a nearly 12% loss in detected photons, attributable to absorption and diffraction caused by the wire. In contrast, when the wire is placed at $x = 0.725$ mm, the photon count remains essentially unchanged, with $f = 0.9888$, consistent with a dark fringe where destructive interference occurs.

Classical diffraction calculations indicate that when the wire is positioned at $x = 0.725$ mm, the fraction of light diffracted to the incorrect detector is negligible—roughly five photons out of every billion are deflected due to wire diffraction. This means that at this specific wire position, the system behaves almost identically to the case without the wire: photons pass through the beam intersection largely unimpeded. Consequently, when detector B in Fig. 1 registers a photon, it most likely originated from the mirror; similarly, a click at detector A corresponds to a photon from the beam splitter. Because scattering is minimal, momentum conservation provides strong evidence for the photon's path, yielding a path information parameter of approximately $K \approx 1$.

Furthermore, the negligible photon loss observed at $x = 0.725$ mm indicates nearly complete destructive interference at this location, implying a local electric field intensity close to zero, $I_0 \approx 0$. However, in the vicinity of this position, Fig. 2 shows a measurable field intensity, $I \neq 0$. As a result, the visibility in this region, defined as

$$V = \frac{I - I_0}{I + I_0} \approx 1, \quad (3)$$

approaches unity. This leads to a measured violation of the complementarity inequality in Eq. 2, producing a clearly paradoxical outcome.

Solutions to this paradox have been proposed for wire grids placed at the dark fringes of the interference pattern, but these explanations have been challenged [7–10]. Importantly, none of the existing proposals address the single-wire case analyzed here. We argue that the paradox reveals a limitation of conventional semi-classical or purely wave-based treatments of interference and diffraction. The difficulty seems to stem from the passive role attributed to the wire: in classical theory, interference is assumed to form independently of the wire's presence.

To resolve the complementarity paradox, we propose that the wire plays a key role in the formation of the interference pattern in the experiment described in Fig. 1. In our interpretation, if the wire is absent, a photon in either beam travels undisturbed from source to detector, even when crossing regions of constructive or destructive interference. However, when the wire is present, the photon interacts with it, and the interference pattern is established.

For the wire to be an active participant, photons must interact with the charges that constitute it. This motivates us to consider photon–charge interactions within the framework of quantum electrodynamics (QED). In QED, all interactions are mediated by virtual particles, which, like real particles, can transfer energy, momentum, angular momentum, and charge. Real particles obey the relativistic energy–momentum relation $E^2 = p^2c^2 + m^2c^4$ and exist as free states, whereas virtual particles do not satisfy this relation, are short-lived, and cannot be isolated.

Due to the fundamental symmetry between light and matter in quantum theory, repeating this experiment with electrons instead of photons would produce a similar contradiction. Electron scattering from a wire treated as a static potential is presented in the Appendix. Photon scattering from a static potential, however, is considerably more complex. In QED, this process is interpreted as *light-by-light* scattering, a fundamentally nonlinear phenomenon [11,12]. Unlike electron scattering, which can often be described using a single leading-order Feynman diagram, photon scattering involves a more intricate set of contributions. The lowest-order process consists of six one-loop Feynman diagrams [11–13]. One representative diagram is shown in Fig. 3, where a virtual electron loop connects the incoming and outgoing photons.

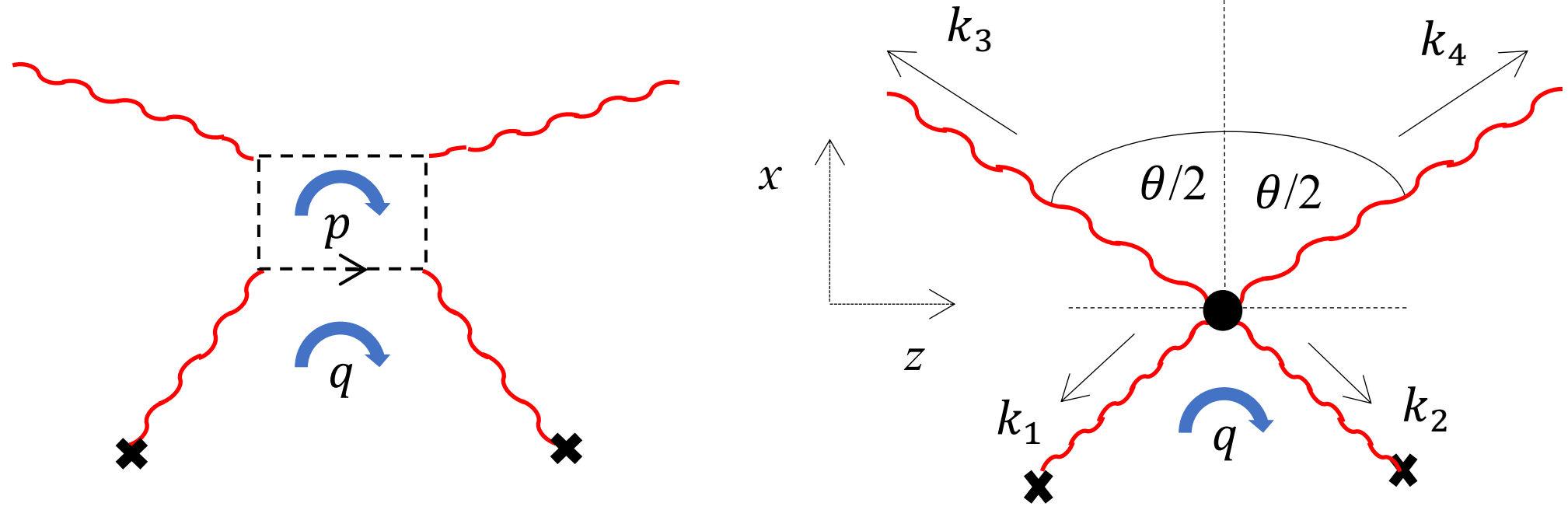

**Fig. 3 | Feynman diagram for photon scattering from a static potential representing a wire.** This diagram illustrates the QED process by which a photon scatters from a static potential, such as a wire. The interaction is mediated by a virtual electron loop, which couples the incoming and

outgoing photons. The incoming photon carries 4-momentum $k_3$, while the scattered photon has 4-momentum $k_4$, with the scattering angle defined as $\theta$. The "x" at the end of the external photon line marks the interaction with the static potential. The wire is modeled as an extended object aligned along the $y$-axis. Momentum is exchanged between the virtual electron loop and the wire via two virtual photons with 4-momenta $k_1$ and $k_2$.

In QED, the process shown in Fig. 3 is closely related to Delbrück scattering, first observed experimentally in 1973 [14]. In Delbrück scattering, a photon is scattered by the Coulomb potential of an atomic nucleus via a virtual electron loop. This process serves as a well-known example of light-by-light scattering in the presence of a static potential. At low photon energies, the amplitude for Delbrück scattering has been derived analytically [13,15]. In this work, we adopt the results presented in Ref. 13, where the theoretical treatment of photon scattering is formulated in a way general enough to accommodate a variety of external potentials beyond the Coulomb field. This generalization allows us to apply the formalism to model photon scattering from the potential associated with a wire, as required for our analysis.

Using the notation in Ref. 13, in the coordinate system in Eq. (31) momentum of the virtual photons is

$$(k_1)^2 = q^2 - 2qk\sin\left(\frac{\theta}{2}\right)\sin\phi + k^2\sin^2\left(\frac{\theta}{2}\right)$$

$$(k_2)^2 = q^2 + 2qk\sin\left(\frac{\theta}{2}\right)\sin\phi + k^2\sin^2\left(\frac{\theta}{2}\right)$$

where $\theta$ is the scattering angle, $k$ is the incoming and outgoing photon energy (in units of electron rest mass), $q$ and $\phi$ are integration variables. After a long calculation, we obtain the leading order contribution to the amplitude for scattering low-energy circular polarized photons from a static potential with cylindrical symmetry. The amplitudes are:

$$\mathrm{M}_{++} = \mathrm{M}_{--} = \iint_{q=0,\phi=0}^{q=\infty,\phi=2\pi} \frac{11}{45}[(\cos\theta - \cos 2\phi)k^2q^2]V(\mu_1)V(\mu_2)q\,dq\,d\phi, \quad (4)$$

$$\mathrm{M}_{+-} = \mathrm{M}_{-+} = \iint_{q=0,\phi=0}^{q=\infty,\phi=2\pi} \frac{1}{45}[(8 - 5\cos\theta - 3\cos 2\phi)k^2q^2]V(\mu_1)V(\mu_2)q\,dq\,d\phi \quad (5)$$

where the sign $+$ $(-)$ stands for right (left) circular polarized light. The Fourier transformed of the potential is $V(\mu_i)$, where $\mu_i = (k_i)^2/4$.

We find that the cylindrical potential barrier used for modeling the wire does not yield accurate results when applied to photon scattering. To address this issue, we introduce the smoother and more physically realistic potential in Fig. 4 to represent the wire. Specifically, we model the wire potential as the convolution of a unit box function and a Gaussian function of effective radius $r = R/8$, where $R$ is the physical radius of the wire. The Fourier transform of this smoothed potential is

$$V(\mu_\mathrm{i}) = (\pi H R^3/8)\,{}_0F_1[2, -\mu_\mathrm{i}R^2]e^{-4\pi^2\mu_\mathrm{i}r^2}, \quad (6)$$

where ${}_0F_1$ is a hypergeometric generalized function. As illustrated in Fig. 4, the modified potential in Eq. 6, yields reasonable and consistent results for both electron and photon scattering cases.

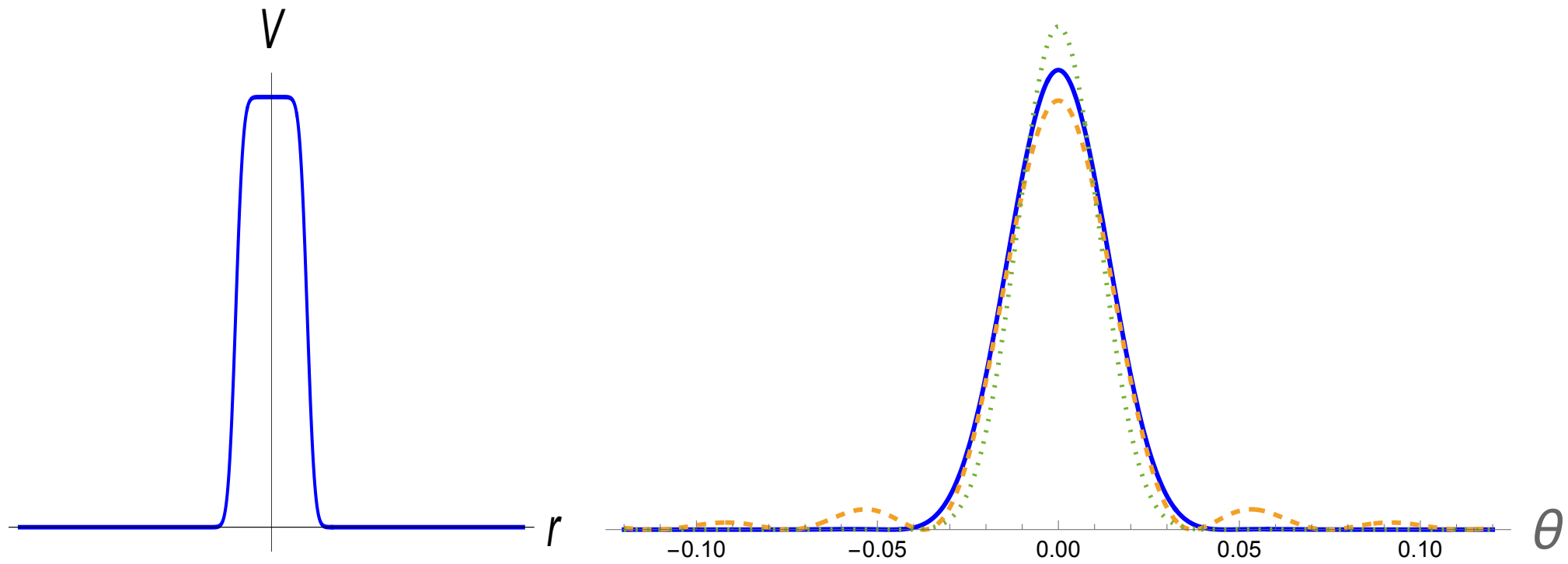

**FIG. 4 | Modified potential barrier.**
The modified potential barrier is smoother than the cylindrical barrier. Fraunhofer diffraction, shown as a dashed line, is plotted as a function of $\theta$. The electron probability amplitude corresponding to the modified potential is shown as a dotted line. The photon probability amplitude, obtained from Eq. 4, is represented by the solid line. For comparison, the area under all three distributions has been normalized to one.

We observe that incorporating diffraction calculations based on QED has not yet resolved the paradox of the apparent violation of the complementarity inequality. To address this, we introduce ladder diagrams, a well-established tool in particle physics for describing bound states and repeated virtual-particle exchanges [16,17]. Consider a particle moving directly toward a wire positioned at the center of a dark fringe. Virtual particles mediate momentum transfers from the wire to the incoming particle. As a result, the particle receives transverse kicks that cause it to orbit around the wire and avoid direct interaction, as illustrated in Fig. 5. Through this redirection, the particle contributes to the formation of the dark fringe, thereby enhancing the interference pattern and yielding maximum visibility ($V = 1$).

After passing the wire, the particle is presented with two indistinguishable forward paths, each corresponding to the same field intensity. In such a scenario, quantum mechanics dictates that the particle will make a *random choice* between the two paths, a process that destroys path information, resulting in $K = 0$. Consequently, the complementarity inequality $K^2 + V^2 \leq 1$ is preserved.

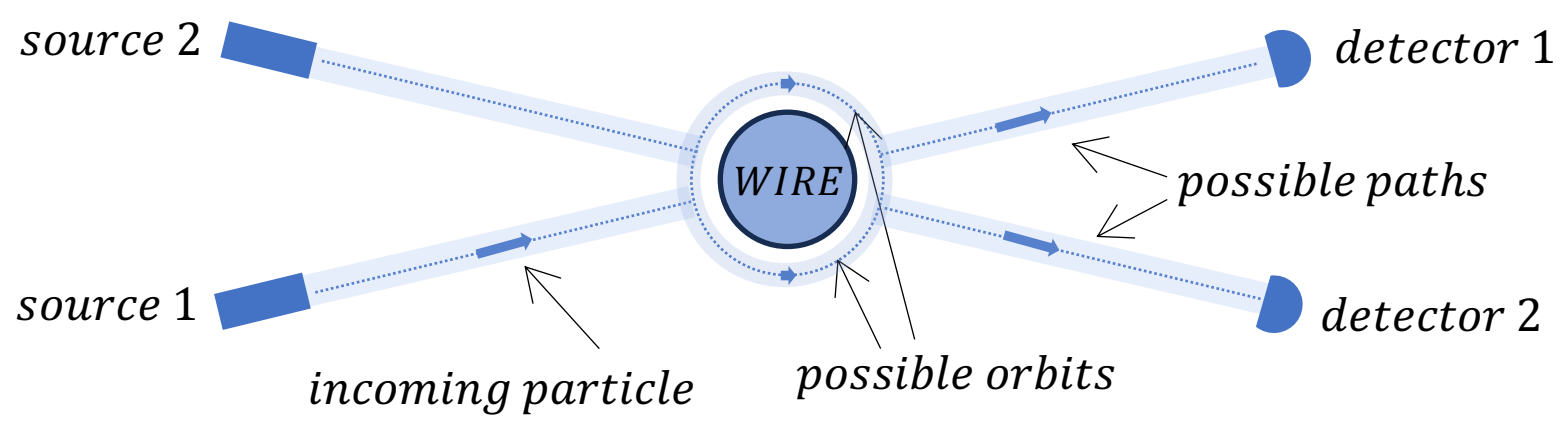

**FIG. 5 | Particle circles the wire and randomly choses of a final path.**
Consider a particle emitted from source 1. As it approaches the wire, the wire can impart momentum to the particle, enabling it to arc around without making direct contact. After passing the wire, the particle faces two equally probable paths, as the wavefunction amplitudes (or field intensities) for both paths are identical. One path leads to detector 1; the other, to detector 2. Note that for the particle to reach detector 2, the wire must again impart momentum to alter the particle's initial trajectory.

We note that the total momentum given to the particle by the surrounding components—the source, wire, and detector—sums to zero, since all these elements are rigidly attached to a common support. As a result, it is fundamentally impossible to determine which specific component transferred momentum to the particle. In addition, the choice of final path is a random process. This uncertainty prevents us from identifying the particle's origin, meaning we cannot tell which source it came from. Consequently, the path information is erased, $K = 0$.

A single Feynman diagram—such as the first one in Fig. 6—is insufficient to explain an electron orbiting the wire. However, an infinite series of diagrams, like those shown in Fig. 6, can collectively provide the momentum needed for the electron to arc around the wire and avoid a direct collision. This series of ladder diagrams effectively represents the exchange of an infinite number of virtual particles. While each individual virtual particle carries only an infinitesimal amount of momentum, their cumulative effect can be substantial, significantly altering the particle's trajectory. In our scenario, the relevant diagrams take the form of spirals rather than simple ladders, since all the "x" symbols represent interactions with the single wire, as shown in Fig. 6.

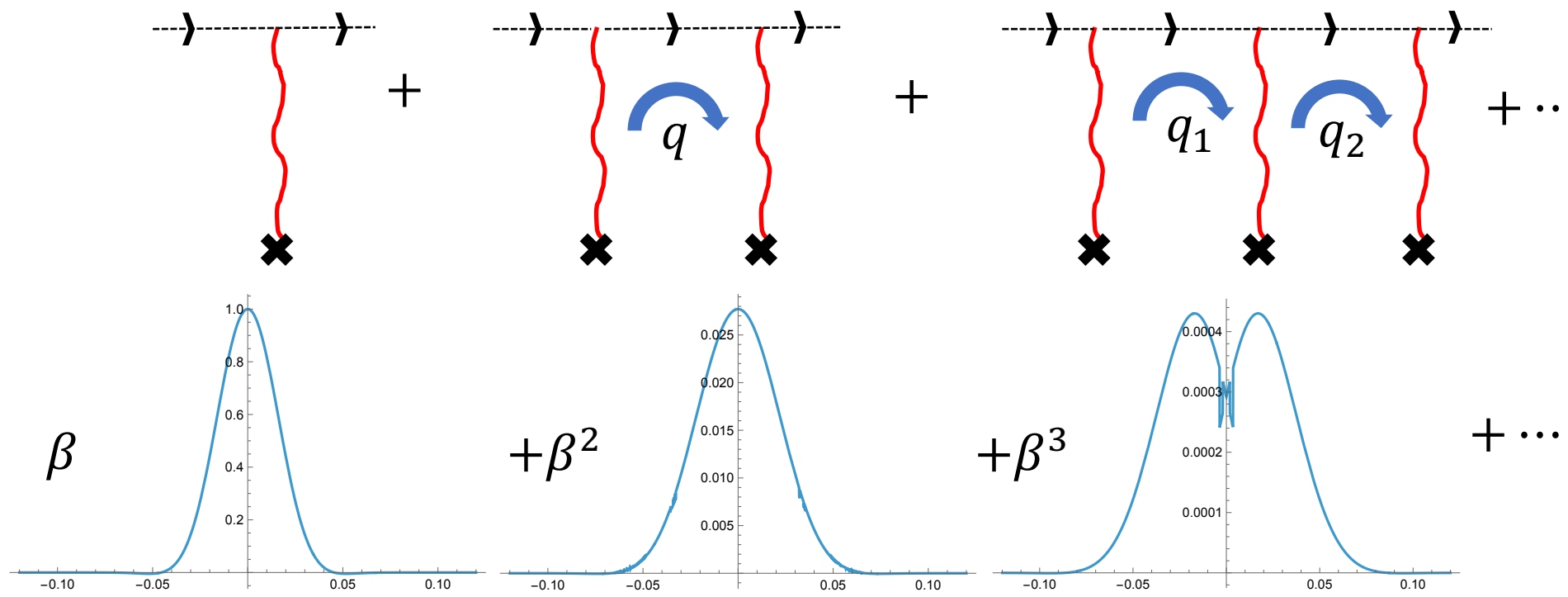

**FIG.** 6 | **Spiral Diagrams for the Electron.**
The first diagram depicts an electron scattering off a potential that represents the wire, indicated by the "x". The interaction strength is governed by a coupling constant $\beta \propto H$, where $H$ characterizes the strength of the wire's field. By choosing $H$ sufficiently large, the contributions from higher-order diagrams become non-negligible. These higher-order diagrams resemble repeated versions of the first and contribute to the transition amplitude $S_{fi}$. Each successive order involves momentum integration over internal lines. In these spiral diagrams, the steps are uncrossed, reflecting the fact that the wire—modeled as an object with infinite mass—does not propagate. It acts purely as a static source of momentum transfer.

Let us consider the second graph in the series in Fig. 6. This graph is formed by two virtual photons interacting with the wire and a virtual electron connecting two basic diagrams. The matrix element is given by

$$S_{fi} = -ie^2 \int d^4x\, d^4y \bar{\psi}_f(x)\gamma^\mu A_\mu(x) S(x-y)\gamma^\nu A_\nu(y)\psi_i(y), \tag{7}$$

where $S$ is the electron propagator. We substitute the wavefunction of the incoming and outgoing electron, replace the potential and obtain

$$S_{fi} = -ie^2\delta(E_f - E_i)\delta(p_f - p_i)_y \int \frac{d^2p}{(2\pi)^2}\bar{u}_f V(p_f - p)\gamma^0 \frac{(E_i\gamma^0 - \vec{p}\cdot\vec{\gamma} + m)}{E_i^2 - p^2 - m^2 + i\epsilon}\gamma^0 V(p - p_i)u_i, \quad (8)$$

There is nothing particularly unusual about the angular integration—it proceeds in the standard way. The integration over the momentum magnitude can be performed using the method of residues. The resulting expression is proportional to $(eH\pi R^3/8)^2$, multiplied by a function that closely resembles the original diffraction pattern shown in Fig. 6. We adjust the height $H$ of the potential to ensure that higher-order diagrams contribute significantly to the scattering amplitude. This is essential, as one of the key requirements [16,17] of the Bethe–Salpeter equation (or the related spectator approximation [17]) is that all relevant diagrams contribute coherently and significantly—enough to generate a pole in the transition matrix element $S_{fi}$

Using the Bethe–Salpeter equation framework, we can identify the key components of the bound-state process: the vertex, the kernel, and the two-particle propagator [16,17]. In our analysis, the vertex is represented by the potential $V(p)$, and the kernel corresponds to the first Feynman diagram in Fig. 6. The two-particle propagator simplifies considerably because the wire, modeled as having infinite mass, does not propagate—only the electron's propagator connects successive kernel interactions. In this infinite-mass limit, the Bethe–Salpeter equation reduces to the so-called spectator (or Gross) equation [17]. Under this approximation, it has been shown that summing an infinite series of diagrams—such as the spiral diagrams in Fig. 6—leads to the Dirac equation with an external potential $V(\vec{r})$ [17]. In essence, the infinite series of spiral diagrams effectively reproduces the exact solution of the Dirac equation for a static potential.

While the Dirac equation offers accurate predictions—for example, matching experimental results for the hydrogen atom—it falls short in capturing finer quantum effects, such as the Lamb shift, which arises from vacuum fluctuations [18]. These effects are accurately explained within QED through virtual particle exchange [19]. Likewise, in systems like positronium, the Dirac equation predicts incorrect energy levels, whereas virtual particle-based approaches yield accurate results. These examples underscore the power of the diagrammatic, field-theoretic approach in capturing higher-order quantum corrections that the Dirac equation, as a single-particle theory, cannot fully describe.

In this context, the exact solution of the Dirac equation with an external potential $V(\vec{r})$, representing the wire, should—at least in principle—capture the cumulative effects of virtual particle exchanges between the electron and the wire. Thus, when two identical, phase-aligned electron beams intersect and interfere while interacting with $V(\vec{r})$, the resulting intensity pattern should match the empirical data shown in Fig. 2. However, although the Dirac equation correctly predicts the final distribution, it obscures the mechanism by which electrons avoid hitting the wire. It does not explicitly reveal the momentum exchanges or local interactions that enforce complementarity. For instance, when the wire is placed at the center of a dark fringe and an electron avoids it, it achieves maximal visibility ($V = 1$); the electron must then randomly choose between two outgoing paths to erase path information($K = 0$). The Dirac equation does not show how the electron receives the large, sudden momentum transfer from the wire necessary to deflect it from its original path.

A similar insight applies to the photon case. Although wave equations predict the interference pattern, it is the virtual particle picture that reveals the internal dynamics of quantum field interactions. The photon case is even more complex than the electron case but follows the same principles. As shown in Fig. 7, the first diagram for photon scattering already involves a momentum integral due to two simultaneous interactions with the wire. In the second diagram, we observe deviations from the original diffraction pattern—similar to what appeared in the third diagram for the electron.

These discrepancies likely arise from our simplified model of the wire as a potential barrier—a rough approximation. In reality, the wire consists of countless point-like charged particles interacting with incoming photons via virtual photon exchange. A more realistic many-body quantum model would capture these interactions more faithfully and likely produce results closer to the first-order diagram. Still, despite its simplicity, our toy model captures the essential physics and provides meaningful insights. Its limitations should not be viewed as flaws, but rather as necessary trade-offs for analytical tractability.

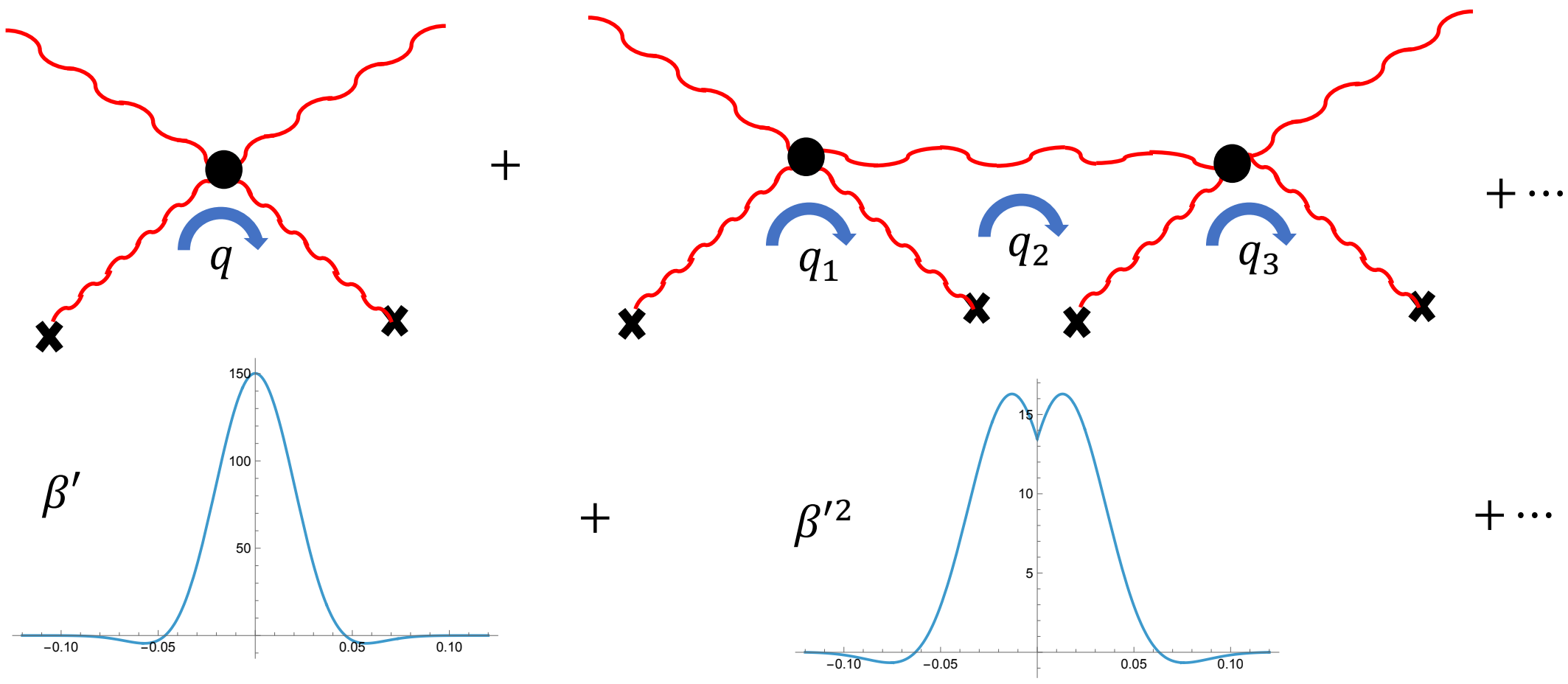

**FIG.** 7 **| Spiral Diagrams for the Photon.**
The first diagram illustrates a photon scattering from a potential that models the wire, represented by the ”x”. The strength of the interaction is governed by a coupling constant $\beta' \propto H^2$, reflecting the fact that photon scattering involves two interactions with the wire. By choosing a sufficiently large value of $H$, the contributions from higher-order diagrams become significant.

Finally, we observe that when the wire is removed in the experiment shown in Fig. 1, there is no longer a source of momentum to transform the photon beam from a uniform flow into one with regions of high and low intensity. In this case, the photon beam continues to propagate uniformly, even though the field values vary across the intersection region. However, when a screen is introduced at the point of beam intersection, the particles within the screen supply the momentum required to deflect the photons, directing them according to the local field values, and an interference pattern becomes visible on the screen.

These considerations suggest a central interpretation: particles do not strictly follow the wavefunction or field. Without external sources of energy and momentum, they behave as free particles, independent of local field configurations. With external sources, virtual particles mediate the required exchanges, enforcing the constraints implied by the wavefunction or field. Thus, the wavefunction or field serves as an informational structure—a blueprint guiding virtual-particle interactions that shape particle trajectories. This interplay links field-theoretic formalism with observed quantum behavior. A clear example is the hydrogen atom, where virtual particles mediate exchanges between electron and proton, enforcing the constraints given by the wavefunction derived from the Dirac equation.

As an application of our findings, consider neutrinos bound to the galaxy. Neutrinos interact via the weak force, with a range of about $10^{-18}$ meters, and via gravity, which is long-range. The wavefunction of Dirac neutrinos orbiting the galaxy is governed by the Dirac equation in curved spacetime. For such neutrinos to exhibit quantum behavior, their wavefunctions must be realized through interactions mediated by virtual particles, much like electrons in atoms interact via virtual photons. In the galactic environment, gravity is the only long-range force acting on neutrinos. Thus, if galactic neutrino distribution displays quantum behavior, it would imply that gravitational interactions involve the exchange of virtual gravitons, supporting the view that gravity is a quantum force. Conversely, if gravity is purely classical, as described by general relativity, galactic neutrinos would behave as a classical gas in a nearly spherical gravitational field. In that case, one could model neutrinos as dark matter and test whether the results reproduce the observed dark matter distribution.

## ACKNOWLEDGMENTS

The authors acknowledge Prof. Michael Lim and Jeffrey Scaturro for their insightful comments and for performing the inspiring experiment. We also thank Leo Han for constructive questions that contributed to this study.

APPENDIX

Electron scattering from a static potential

To describe the interaction of an electron with a wire, we model the wire as a static cylindrical potential barrier. Within the QED framework, this interaction occurs via the exchange of virtual photons between the electron and the external source—here, the electromagnetic field associated with the wire. For an electron interacting with an external potential, the leading-order process is represented by the Feynman diagram shown in Fig. 8. The corresponding scattering amplitude is given by:

$$S_{fi} = -ie \int d^4x\, \bar{\psi}_f(x)\gamma^\mu A_\mu(x)\psi_i(x), \tag{9}$$

where $\psi_i = au(p_i, s_i)e^{-ip_i \cdot x}$ represents an incoming electron, $a$ is a normalization constant, $p$ is the 4-momentum, $s$ is the spin, $u$ is a spinor and $x$ is the space-time coordinate [19]. The outgoing electron is represented by $\bar{\psi}_f = a\bar{u}(p_f, s_f)e^{ip_f \cdot x}$. The vector potential $A_\mu$ is a virtual photon due to a static potential, $A_\mu = (V, 0,0,0)$.

The wire is long, perpendicular to the $xz$-plane; thus, we model the wire as a cylindrical potential which in polar coordinates is given by $V = H,\ 0 \leq s \leq R$, and zero otherwise. The dot product may be written as $(p_f - p_i) \cdot x = (E_f - E_i)t - qs\cos\phi$, where $q$ is the magnitude of $\vec{q} = (\vec{p}_f - \vec{p}_i)_{xz}$ and $\phi$ is the angle between $\vec{q}$ and the position vector. Energy conservation requires that $E_f = E_i$. We note that $|\bar{u}\gamma^0 u|^2$ for forward scattering is significant only if the spin of the electron does not flip. Evaluating Eq. 1 at $p_{ix} = 0$, $p_{iz} = p,\ p_{fz} = p\cos\theta$, $p_{fx} = p\sin\theta$ and in the limit $p/c \ll m$, results in an amplitude proportional to the Fourier transformed of the potential

$$S_{fi} \propto V(p) = (\pi H R^2)\,{}_0F_1\left[2, -R^2p^2\sin^2\left(\frac{\theta}{2}\right)\right], \tag{10}$$

where ${}_0F_1$ is a hypergeometric generalized function.

According to Babinet's principle, diffraction by a slit and by a wire are mathematically complementary problems [1]. As such, the diffraction pattern produced by a thin wire can be approximated using the diffraction from a complementary slit of the same width. In classical optics, Fraunhofer diffraction from a narrow slit is described by the intensity pattern [1]:

$$I \propto \mathrm{sinc}^2(Rp\sin\theta). \tag{11}$$

In Fig. 8, we compare the quantum mechanical probability amplitude for an electron scattering off a potential representing the wire, with the classical Fraunhofer diffraction pattern of a slit. While both approaches exhibit qualitatively similar features such as central maxima and side lobes, there are key differences. In the quantum approach, the wire is treated as a three-dimensional potential structure, whereas in classical Fraunhofer theory the wire is approximated as a two-dimensional object—an infinitely long, thin screen. This dimensional discrepancy has been noted in the literature as a potential source of systematic overestimation of the wire's effective radius when applying classical Fraunhofer diffraction to interpret experimental data [20,21]. The quantum treatment, by contrast, offers a more accurate framework for analyzing the interaction.

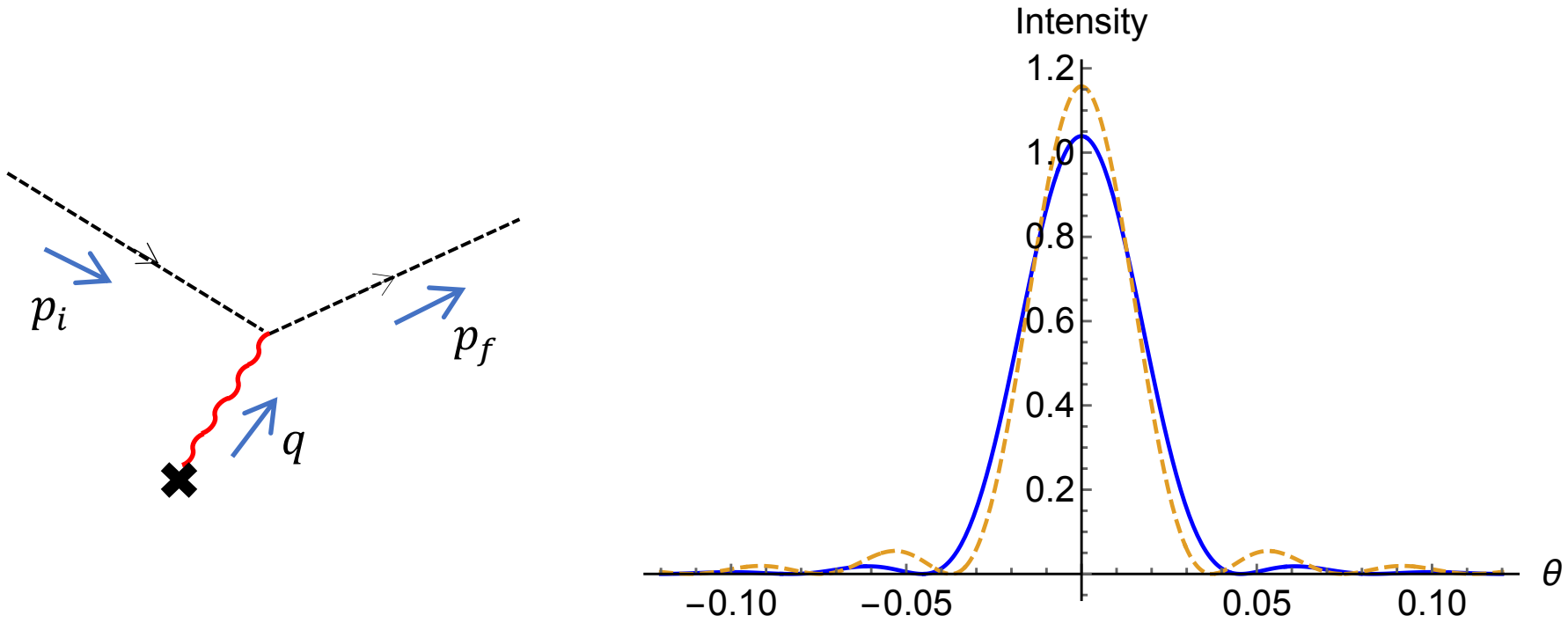

**Fig. 8 | Comparison of quantum and classical diffraction from a wire.**
The Feynman diagram on the left represents an electron with a wavelength of $633 \times 10^{-9}$ nm scattering from a static potential barrier that models a wire of diameter 17 $\mu$m. On the right, the solid blue curve represents the probability amplitude calculated using QED. The dashed curve shows the corresponding Fraunhofer diffraction pattern. To enable comparison, the area under each curve has been normalized to be equal. Notably, the location of the first minimum in the QED result is shifted outward relative to the classical prediction, indicating that the Fraunhofer approach tends to overestimate the effective wire diameter compared to the more accurate quantum model.